\documentclass[11pt,a4paper]{article}
\usepackage[utf8]{inputenc}
\usepackage[T1]{fontenc}
\usepackage[british]{babel}
\usepackage[sc,osf]{mathpazo}
\usepackage[margin=2.6cm]{geometry}
\usepackage{amsmath,amssymb,amsthm,mathtools}
\usepackage[numbers,sort&compress]{natbib}
\usepackage{microtype}
\usepackage{booktabs}
\usepackage{enumitem}
\usepackage{authblk}
\usepackage[colorlinks=true,citecolor=blue,urlcolor=blue,linkcolor=blue]{hyperref}

\theoremstyle{plain}
\newtheorem{theorem}{Theorem}[section]

\newtheorem{proposition}[theorem]{Proposition}

\theoremstyle{definition}

\newcommand{\Tr}{\mathrm{Tr}}

\DeclareMathOperator{\diag}{diag}
\DeclareMathOperator{\artanh}{artanh}

\hypersetup{pdftitle={Sharp Purity Based Entropy Bounds and Machine Learning Assisted Entropy Estimation},
            pdfauthor={Komal Kumar,Shree Prakash}   }

\title{\textbf{Sharp Purity Based Entropy Bounds and Machine Learning Assisted Entropy Estimation}}

\author[1]{Komal Kumar\thanks{komal.kumar@vit.ac.in}}
\author[2]{Shree Prakash\thanks{shree.prakash@vit.ac.in}}
\affil[1]{\small Department of Mathematics,
School of Advanced Sciences, Vellore Institute of Technology,
Chennai, Tamil Nadu-600127, India}
\affil[2]{\small School of Computer Science and Engineering, Vellore Institute of Technology, Chennai, Tamil Nadu-600127, India}
\date{}

\begin{document}

\maketitle
\begin{abstract}
Purity quantifies quantum-state mixedness, motivating the study of its role in controlling entropy variations. A sharp purity-based continuity bound is established for the von Neumann entropy of qubit states. The bound is attainable for every admissible purity difference and strictly improves the existing purity bound. In higher dimensions, an explicit sharp bound is derived within the family of pure states mixed with the maximally mixed state. Further continuity estimates are obtained for conditional entropy, quantum mutual information, quantum conditional mutual information, and squashed entanglement within specified bipartite and tripartite
families. An application to machine learning assisted qubit entropy estimation illustrates how the sharp bound converts purity estimation errors into rigorous entropy error bounds.
\end{abstract}

\section{Introduction}\label{intro}

Quantum entropies and related information measures
play a central role in quantum information processing~\cite{Zoller2005}.
They characterize the uncertainty, correlations,
and entanglement associated with quantum states
and provide mathematical tools for studying
information-processing tasks.
Important examples include the von Neumann
entropy~\cite{Wilde2017, vonNeumann1955}, conditional von Neumann
entropy~\cite{CerfAdami1997, CerfAdami1999}, relative entropy-based
quantities~\cite{Vedral2002}, quantum mutual information
(QMI), quantum conditional mutual information
(QCMI)~\cite{NielsenChuang2010}, and squashed
entanglement~\cite{ChristandlWinter2004}.
Understanding the stability of these quantities
under changes in the underlying state is therefore
of both mathematical and operational interest.

Continuity bounds quantify this stability by
controlling the change in an information measure
in terms of a prescribed measure of closeness
between states. In finite-dimensional systems,
such bounds are commonly expressed using the
trace distance. The bounds of Fannes and
Audenaert~\cite{Fannes1973,Audenaert2007} for the von Neumann entropy
and of Alicki and Fannes~\cite{AlickiFannes2004} for the
conditional von Neumann entropy provide
fundamental tools in this setting.
Subsequent developments have refined these bounds
and extended continuity estimates to other
quantum information
measures~\cite{ChristandlWinter2004,Fannes1973,Audenaert2007,AlickiFannes2004,AudenaertEisert2005,Shirokov2010,Shirokov2016,LiWinter2014}.
In recent years, research works have been carried to further tighten the uniform continuity bounds~\cite{Winter2016,Shirokov2017,HansonDatta2017}.
Also, in \cite{HoYeung2010,Zhang2007}, authors have discussed various proofs of tight uniform continuity bounds of Shannon and von Neumann entropies, and tight local continuity bounds for von Neumann entropy has been given in\cite{HansonDatta2018}. For a detailed review, one may see \cite{Shirokov2023}.

% In \cite{RehmanShin2019}, author have gave continuity bounds of von Neumann entropy based on purity difference of two state. Then, we derive new intuitive continuity bounds based on purity difference, which are significantly tighter than the aforementioned bounds.

% The purity of state is given as
% $P(\rho)=\operatorname{Tr}(\rho^2)$. For two density operators $\rho$ and $\sigma$,
% we consider the absolute purity difference
% \[
%     \epsilon=
%     \left|
%         \operatorname{Tr}(\rho^2)
%         -\operatorname{Tr}(\sigma^2)
%     \right|.
% \]
% Although this quantity is symmetric, nonnegative,
% and satisfies the triangle inequality, it is
% not a metric on the state space: distinct states
% can have equal purity. In qubit setting, authors established the continuity bounds for von Neumann entropy \cite{RehmanShin2019}, given as
% \begin{equation}{\label{rs}}
%     |S(\rho)-S(\sigma)|
%     \leq h\left(\sqrt{\frac{\epsilon}{2}}\right),
% \end{equation}
% where $h$ denotes the binary entropy.
% In the present work, we establish the sharp bound for von Neumann for qubit 
% \begin{equation}{\label{pb}}
%     |S(\rho)-S(\sigma)|
%     \leq
%     h\left(
%         \frac{1-\sqrt{1-2\epsilon}}{2}
%     \right),
%     \qquad
%     0\leq\epsilon\leq\frac12.
% \end{equation}

In \cite{RehmanShin2019}, the authors established continuity bounds for the von Neumann entropy in terms of the purity difference between two quantum states. In the present work, we derive new, intuitive continuity bounds based on the purity difference that are significantly tighter than these existing bounds.

The purity of a quantum state $\rho$ is defined as
$P(\rho)=\operatorname{Tr}(\rho^2)$. For two density operators $\rho$ and $\sigma$, we consider the absolute purity difference
\[
    \epsilon=
    \left|
        \operatorname{Tr}(\rho^2)
        -\operatorname{Tr}(\sigma^2)
    \right|.
\]
Although this quantity is symmetric, nonnegative, and satisfies the triangle inequality, it is not a metric on the state space, since distinct states can have equal purity.

For qubit states, the authors of \cite{RehmanShin2019} established the following continuity bound for the von Neumann entropy:
\begin{equation}\label{rs}
    |S(\rho)-S(\sigma)|
    \leq h\left(\sqrt{\frac{\epsilon}{2}}\right),
\end{equation}
where $h$ denotes the binary entropy function. In the present work, we establish the following sharp bound for the von Neumann entropy of qubit states:
\begin{equation}\label{pb}
    |S(\rho)-S(\sigma)|
    \leq
    h\left(
        \frac{1-\sqrt{1-2\epsilon}}{2}
    \right),
    \qquad
    0\leq\epsilon\leq\frac12.
\end{equation}

The bound is optimal as a function of the purity difference alone: for every admissible $\epsilon$, there exist qubit states attaining equality. It strictly improves the bound given by Equation(\ref{rs}) for $0<\epsilon<\tfrac12$, while the two bounds coincide at the endpoints.

In higher dimensions, we restrict our discussion to the family of states
\[
    \rho_t=
    t|\psi\rangle\langle\psi|
    +(1-t)\frac{I_d}{d},
    \qquad 0\leq t\leq1,
\]
where $|\psi\rangle$ is a fixed normalized pure state. Within this family, the purity is given by
\[
    \operatorname{Tr}(\rho_t^2)
    =\frac{1+(d-1)t^2}{d}.
\]
Thus, the purity uniquely determines the parameter $t$ and, consequently, the spectrum of $\rho_t$. We exploit this relation to express the entropy as a function of purity and subsequently derive continuity bounds for various entropic quantities.

We extend the entropy bound to specified bipartite and tripartite families. For isotropic states with nonnegative mixing parameter, maximally mixed marginals yield sharp continuity bounds for conditional von Neumann entropy, QMI, coherent information, and the relative entropy-based quantities considered here. We also derive a purity-based continuity bound for squashed entanglement within this family. For generalized GHZ states mixed with the maximally mixed state, we obtain a purity-based QCMI bound. This conclusion is restricted to the specified family and reference state. Finally, we illustrate an application of the sharp qubit continuity bound to machine-learning-assisted entropy estimation.

The manuscript is organized as follows. In Sec.~\ref{intro}, we introduce the required definitions and discuss the limitations of purity difference as a measure of state closeness. In Sec.~\ref{sec:qubit-purity-bound}, we prove the sharp qubit entropy bound.
In Sec.~\ref{sec:restricted-families}, we derive bounds
for von Neumann entropy within specified higher-dimensional state families, and obtained various entropic bounds in bipartite and tripartite system for some particular state. In Sec.~\ref{sec:ann-application}, we investigate an application to machine-learning-assisted entropy estimation. Finally, we conclude in Sec.~\ref{conclusion}.

The manuscript is organized as follows. In Sec.~\ref{intro}, we introduce the required definitions and discuss the properties of purity difference and its role in quantifying entropy variations. In Sec.~\ref{sec:qubit-purity-bound}, we prove the sharp qubit entropy bound. In Sec.~\ref{sec:restricted-families}, we derive bounds for the von Neumann entropy within specified higher-dimensional state families and establish bounds for various entropic quantities in particular bipartite and tripartite state families. In Sec.~\ref{sec:ann-application}, we investigate an application to machine-learning-assisted entropy estimation. Finally, we conclude in Sec.~\ref{conclusion}.

%%%%%%%%%%%%%%%%%%%%%%%%%%%%%%%%%%%%%%%%%%%%%%%%%%%%%%%%%%%%%%%%%%%%5

% \section{Preliminaries and notation}\label{sec:prelim}
%============================================================

\section{Entropic Bound for Qubits}
\label{sec:qubit-purity-bound}

\subsection{Bounds on von Neumann entropy}
Let $\mathrm{D}(\mathbb{H}_X)$ denote the set of density operators on the Hilbert space $\mathbb{H}_X$. The von Neumann entropy of a state $\rho\in\mathrm{D}(\mathbb{H}_X)$ is defined as
\[
    S(\rho)=-\Tr(\rho\log_{2}\rho).
\]
The binary Shannon entropy is defined by
\[
    h(x)=-x\log_{2}x-(1-x)\log_{2}(1-x),
    \qquad 0\leq x\leq1,
\]
with the convention $0\log_{2}0=0$. Throughout, $\log$ denotes the logarithm to base $2$, while $\ln$ denotes the natural logarithm. Here, we consider qubit states, so $\dim(X)=2$.

\begin{theorem}[Sharp purity-based entropy bound]
\label{thm:sharp-qubit-bound}
Let $\rho,\sigma\in\mathrm{D}(\mathbb{H}_X)$ be qubit states, and let
\[
    \epsilon
    =\left|\Tr(\rho^{2})-\Tr(\sigma^{2})\right|.
\]
Then, for $0\leq\epsilon\leq\tfrac12$,
\begin{equation}
    \left|S(\rho)-S(\sigma)\right|
    \leq
    h\left(\frac{1-\sqrt{1-2\epsilon}}{2}\right).
    \label{eq:sharp-qubit-bound}
\end{equation}
The bound is optimal: for every $\epsilon\in[0,\tfrac12]$,
there exist qubit states that attain equality.
\end{theorem}

\begin{proof}
Let $\omega\in\mathrm{D}(\mathbb{H}_X)$ have
eigenvalues $\lambda$ and $1-\lambda$, and denote its
purity by $P=\Tr(\omega^{2})$. Since
\[
    P=\lambda^{2}+(1-\lambda)^{2},
\]
its eigenvalues are
\begin{equation}
    \lambda_{\pm}
    =\frac{1\pm\sqrt{2P-1}}{2},
    \qquad \frac12\leq P\leq 1.
    \label{eq:qubit-eigenvalues}
\end{equation}
Consequently, the entropy of a qubit is determined by
its purity:
\begin{equation}
    S(\omega)=s(P),
    \qquad
    s(P):=
    h\left(\frac{1+\sqrt{2P-1}}{2}\right).
    \label{eq:entropy-purity-function}
\end{equation}

We first show that $s$ is decreasing and strictly
concave on $(\tfrac12,1)$. Set
\[
    r=\sqrt{2P-1}\in(0,1).
\]
Using
\[
    h'(x)=\log_{2}\left(\frac{1-x}{x}\right),
    \qquad
    \frac{dr}{dP}=\frac{1}{r},
\]
we obtain
\begin{align}
    s'(P)
    &=
    \frac{1}{2r}
    \log_{2}\left(\frac{1-r}{1+r}\right)
    \nonumber\\
    &=
    -\frac{\artanh r}{r\ln 2}
    \label{eq:first-derivative}
\end{align}
where
\[
    \artanh r
    =\frac12\ln\left(\frac{1+r}{1-r}\right).
\]

We now verify that $s$ is strictly decreasing.
For $\tfrac12<P<1$, we have
$r=\sqrt{2P-1}\in(0,1)$, and hence
\[
    0<1-r<1+r.
\]
Dividing by $1+r>0$ gives
\[
    0<\frac{1-r}{1+r}<1.
\]
Since the base-$2$ logarithm is negative on $(0,1)$,
\[
    \log_2\left(\frac{1-r}{1+r}\right)<0.
\]
Together with $\frac{1}{2r}>0$, this implies
\[
    s'(P)
    =\frac{1}{2r}
      \log_2\left(\frac{1-r}{1+r}\right)
    <0.
\]
Thus $s$ is strictly decreasing on $(\tfrac12,1)$.
By continuity at the endpoints, $s$ is decreasing
on the entire interval $[\tfrac12,1]$.

In particular, whenever
$\tfrac12\leq P\leq P+\epsilon\leq1$, we have
\[
    s(P)\geq s(P+\epsilon).
\]
Therefore, if
$P=\operatorname{Tr}(\sigma^2)$ and
$P+\epsilon=\operatorname{Tr}(\rho^2)$, then
\[
    |S(\rho)-S(\sigma)|
    =|s(P+\epsilon)-s(P)|
    =s(P)-s(P+\epsilon).
\]

Differentiating again gives
\begin{equation}
    s''(P)
    =
    -\frac{1}{r^{3}\ln 2}
    \left(
        \frac{r}{1-r^{2}}-\artanh r
    \right).
    \label{eq:second-derivative}
\end{equation}
For $0<r<1$, we have
\[
    \artanh r
    =
    \int_{0}^{r}\frac{du}{1-u^{2}}
    <
    \int_{0}^{r}\frac{du}{1-r^{2}}
    =
    \frac{r}{1-r^{2}}.
\]
Thus $s''(P)<0$, establishing strict concavity.

Now suppose that $0<\epsilon<\tfrac12$. Without loss
of generality, assume that
\[
    \Tr(\rho^{2})\geq\Tr(\sigma^{2}).
\]
Writing $P=\Tr(\sigma^{2})$, we have
\[
    \Tr(\rho^{2})=P+\epsilon,
    \qquad
    \frac12\leq P\leq 1-\epsilon.
\]
Since $s$ is decreasing,
\begin{equation}
    |S(\rho)-S(\sigma)|
    =s(P)-s(P+\epsilon).
    \label{eq:entropy-difference}
\end{equation}
For fixed $\epsilon$, define
\[
    G_{\epsilon}(P)
    =s(P)-s(P+\epsilon).
\]
On the interior of its domain,
\[
    G_{\epsilon}'(P)
    =s'(P)-s'(P+\epsilon)>0,
\]
because $s'$ is strictly decreasing. By continuity,
$G_{\epsilon}$ attains its maximum at $P=1-\epsilon$.
Therefore,
\begin{align}
    |S(\rho)-S(\sigma)|
    &\leq s(1-\epsilon)-s(1)
    \nonumber\\
    &=h\left(\frac{1+\sqrt{1-2\epsilon}}{2}\right)
    \nonumber\\
    &=h\left(\frac{1-\sqrt{1-2\epsilon}}{2}\right),
    \label{eq:bound-derivation}
\end{align}
where we used $s(1)=0$ and the symmetry
$h(x)=h(1-x)$.

It remains to consider the endpoints. If $\epsilon=0$,
the states have equal purity and hence the same
eigenvalue multiset by Eq.~\eqref{eq:qubit-eigenvalues}.
Their entropies therefore coincide. If
$\epsilon=\tfrac12$, their purities must be $1$ and
$\tfrac12$. Thus one state is pure and the other is
maximally mixed, giving
\[
    |S(\rho)-S(\sigma)|=1=h\left(\frac12\right).
\]
Hence Eq.~\eqref{eq:sharp-qubit-bound} holds throughout
$[0,\tfrac12]$.

Finally, fix $\epsilon\in[0,\tfrac12]$ and define
\[
    q_{\epsilon}
    =\frac{1-\sqrt{1-2\epsilon}}{2}.
\]
Consider the states
\begin{equation}
    \rho_{\epsilon}
    =
    \begin{pmatrix}
        1 & 0\\
        0 & 0
    \end{pmatrix},
    \qquad
    \sigma_{\epsilon}
    =
    \begin{pmatrix}
        1-q_{\epsilon} & 0\\
        0 & q_{\epsilon}
    \end{pmatrix}.
    \label{eq:saturating-states}
\end{equation}
Since $2q_{\epsilon}(1-q_{\epsilon})=\epsilon$,
\[
    \Tr(\rho_{\epsilon}^{2})=1,
    \qquad
    \Tr(\sigma_{\epsilon}^{2})
    =(1-q_{\epsilon})^{2}+q_{\epsilon}^{2}
    =1-\epsilon.
\]
Moreover,
\[
    |S(\rho_{\epsilon})-S(\sigma_{\epsilon})|
    =h(q_{\epsilon})
    =h\left(\frac{1-\sqrt{1-2\epsilon}}{2}\right).
\]
Thus equality is attained for every admissible
$\epsilon$. No smaller function of $\epsilon$ alone
can bound the entropy difference for all qubit states.
\end{proof}

% \begin{remark}[Relation to earlier entropy--purity results]
% The relationship between entropy and purity is part
% of the classical theory of entropy and the index of
% coincidence. In particular, the binary entropy--purity
% curve and its concavity are discussed by
% Harremo\"es and Tops{\o}e~\cite{HarremoesTopsoe2001}.
% The proof above explicitly derives the associated
% optimal two-state bound.
% \end{remark}

\subsection{Comparison with the existing purity-based bound}
\label{subsec:comparison-rehman}

For qubit states $\rho$ and $\sigma$, let
\[
    \epsilon
    =\left|\Tr(\rho^{2})-\Tr(\sigma^{2})\right|
    \in\left[0,\frac12\right].
\]
Authors in~\cite{RehmanShin2019} established
\begin{equation}
    |S(\rho)-S(\sigma)|
    \leq R(\epsilon),
    \qquad
    R(\epsilon):=
    h\left(\sqrt{\frac{\epsilon}{2}}\right).
    \label{eq:rehman-bound}
\end{equation}
The bounds in Eq.~\eqref{eq:sharp-qubit-bound} is denoted by
\[
    B(\epsilon):=
    h\left(\frac{1-\sqrt{1-2\epsilon}}{2}\right).
\]

\begin{proposition}
\label{prop:comparison-rehman}
For every $\epsilon\in[0,\tfrac12]$,
\begin{equation}
    B(\epsilon)\leq R(\epsilon),
    \label{eq:bound-comparison}
\end{equation}
with equality if and only if $\epsilon=0$ or $\epsilon=\tfrac12$.
\end{proposition}

\begin{proof}
Let
\[
    q=\frac{1-\sqrt{1-2\epsilon}}{2}.
\]
Then $q\in[0,\tfrac12]$ and
\[
    \epsilon=2q(1-q).
\]
Therefore,
\[
    B(\epsilon)=h(q),
    \qquad
    R(\epsilon)=h\left(\sqrt{q(1-q)}\right).
\]
Since $q\leq\tfrac12$, we have
\[
    q^{2}\leq q(1-q)\leq\frac14,
\]
and hence
\[
    0\leq q\leq\sqrt{q(1-q)}\leq\frac12.
\]
The binary entropy is strictly increasing on
$[0,\tfrac12]$, as follows from
\[
    h'(x)=\log_{2}\left(\frac{1-x}{x}\right)>0,
    \qquad 0<x<\frac12.
\]
It follows that
\[
    B(\epsilon)=h(q)
    \leq h\left(\sqrt{q(1-q)}\right)
    =R(\epsilon).
\]
By strict monotonicity, equality holds precisely when
\[
    q=\sqrt{q(1-q)}.
\]
Squaring both sides gives
\[
    q(2q-1)=0.
\]
Thus $q=0$ or $q=\tfrac12$, corresponding respectively
to $\epsilon=0$ and $\epsilon=\tfrac12$.
For $0<\epsilon<\tfrac12$, the inequality is strict.
\end{proof}

Now, from Theorem~\ref{thm:sharp-qubit-bound} and
Proposition~\ref{prop:comparison-rehman}, we obtain
\begin{equation}
    |S(\rho)-S(\sigma)|
    \leq B(\epsilon)
    \leq R(\epsilon).
    \label{eq:combined-comparison}
\end{equation}
Thus, the sharp bound implies the existing bound given in \cite{RehmanShin2019} for the entire admissible range.

\paragraph{Equality and attainability.}
The equality $B(\epsilon)=R(\epsilon)$ concerns the
comparison of the two bounding functions.
It should be distinguished from saturation of the
entropy inequality
\[
    |S(\rho)-S(\sigma)|=B(\epsilon).
\]
The latter is possible for every
$\epsilon\in[0,\tfrac12]$, as demonstrated by the
states in Eq.~\eqref{eq:saturating-states}.

More precisely, for $0<\epsilon<\tfrac12$, equality
in the sharp bound holds if and only if the larger
of the two purities equals $1$. Indeed, the function
$G_{\epsilon}$ in the proof of
Theorem~\ref{thm:sharp-qubit-bound} is strictly
increasing, so its maximum is attained only when
the smaller purity is $1-\epsilon$.
Equivalently, one state must be pure and the other
must have purity $1-\epsilon$. Their eigenvectors
need not coincide.

In contrast, for every pair with
$0<\epsilon<\tfrac12$,
\[
    |S(\rho)-S(\sigma)|
    \leq B(\epsilon)<R(\epsilon),
\]
so the bound in \cite{RehmanShin2019} cannot be saturated at an interior value of $\epsilon$.

At $\epsilon=0$, both bounds equal zero, and equal
qubit purities imply equal entropies. At
$\epsilon=\tfrac12$, the states are necessarily
pure and maximally mixed, respectively, so
\[
    |S(\rho)-S(\sigma)|
    =B\left(\frac12\right)
    =R\left(\frac12\right)
    =1.
\]

\paragraph{Example.}
Consider
\[
    \rho=\diag(1,0),
    \qquad
    \sigma=\diag\left(\frac9{10},\frac1{10}\right).
\]
Their purity difference is
\[
    \epsilon
    =1-\left(\frac{81}{100}+\frac1{100}\right)
    =\frac9{50}.
\]
The actual entropy difference and the sharp bound
coincide:
\[
    |S(\rho)-S(\sigma)|
    =B\left(\frac9{50}\right)
    =h\left(\frac1{10}\right)
    \approx 0.468996.
\]
However, bound given in \cite{RehmanShin2019} gives
\[
    R\left(\frac9{50}\right)
    =h\left(\frac3{10}\right)
    \approx 0.881291.
\]
This example demonstrates both the attainability of the sharp bound and its strict improvement over the existing bound.

\section{Purity-based continuity bounds for depolarized
pure-state families}
\label{sec:restricted-families}

The absolute difference of purities does not, in general,
control entropy differences in dimensions greater than two.
We therefore restrict attention to families for which the
purity determines the mixing parameter. Throughout this
subsection, all logarithms are taken to base~$2$, and
\[
h(x)=-x\log_2 x-(1-x)\log_2(1-x),
\]
with the convention $0\log_2 0=0$.

Fix a normalized vector $|\psi\rangle\in\mathbb{H}_X$,
where $\text{dim}(X)=d\geq 2$, and consider
\begin{equation}
\rho_t
=
t|\psi\rangle\langle\psi|
+
(1-t)\frac{I_d}{d},
\qquad 0\leq t\leq 1.
\label{eq:depolarized-family}
\end{equation}
The eigenvalues of $\rho_t$ are
\[
\lambda_1(t)=\frac{1+(d-1)t}{d},
\qquad
\lambda_2(t)=\cdots=\lambda_d(t)=\frac{1-t}{d}.
\]
Consequently,
\begin{equation}
P_d(t):=\operatorname{Tr}(\rho_t^2)
=\frac{1+(d-1)t^2}{d},
\qquad
S(\rho_t)=F_d(t),
\label{eq:family-purity-entropy}
\end{equation}
where
\begin{equation}
F_d(t)
=
h\left(\frac{(d-1)(1-t)}{d}\right)
+
\frac{(d-1)(1-t)}{d}\log_2(d-1).
\label{eq:Fd-definition}
\end{equation}

For $0\leq\epsilon\leq(d-1)/d$, define
\begin{align}
a_d(\epsilon)
&=\sqrt{\frac{d\epsilon}{d-1}},
&
b_d(\epsilon)
&=\sqrt{1-\frac{d\epsilon}{d-1}},
\label{eq:ab-definition}\\
B_d(\epsilon)
&=
\max\left\{
\log_2d-F_d\bigl(a_d(\epsilon)\bigr),
F_d\bigl(b_d(\epsilon)\bigr)
\right\}.
\label{eq:Bd-definition}
\end{align}
In particular, $B_d(0)=0$, and $B_d$ is continuous.

\begin{theorem}[Sharp entropy bound within the family]
\label{thm:family-entropy}
Let $\rho_t$ and $\rho_u$ belong to
Eq.~\eqref{eq:depolarized-family}, and let
\[
\epsilon
=
\left|
\operatorname{Tr}(\rho_t^2)
-
\operatorname{Tr}(\rho_u^2)
\right|~,~~ \epsilon \in [0,\frac{d-1}{d}]
\]
Then
\begin{equation}
|S(\rho_t)-S(\rho_u)|
\leq B_d(\epsilon).
\label{eq:family-entropy-bound}
\end{equation}
For every admissible $\epsilon$, equality is attained
by a pair of states in this family.
\end{theorem}

\begin{proof}
Throughout, $d\ge 2$ and the parameters of the family
satisfy $0\le t,u\le 1$. Recall that
\[
P_d(t)=\operatorname{Tr}(\rho_t^2)
=\frac{1}{d}+\frac{d-1}{d}t^2.
\]
Thus the entropy can be expressed as a function of purity:
\[
s_d(P)
=
F_d\!\left(\sqrt{\frac{dP-1}{d-1}}\right),
\qquad \frac{1}{d}\le P\le 1.
\]
This function is continuous on $[1/d,1]$, with
\[
s_d(1/d)=\log_2 d,
\qquad s_d(1)=0.
\]

\medskip
\noindent\underline{\textbf{Monotonicity of $s_d$.}}
For $1/d<P<1$, set
\[
t=\sqrt{\frac{dP-1}{d-1}}\in(0,1),
\qquad
L(t)=\ln\!\left(\frac{1+(d-1)t}{1-t}\right).
\]
Then
\[
\frac{dt}{dP}=\frac{d}{2(d-1)t}.
\]
Writing
\[
x(t)=\frac{1+(d-1)t}{d},
\qquad
y(t)=\frac{1-t}{d},
\]
we have
\[
F_d(t)
=-x(t)\log_2 x(t)
-(d-1)y(t)\log_2 y(t).
\]
Since $x'(t)=(d-1)/d$ and $y'(t)=-1/d$,
\begin{align*}
F_d'(t)
&=-\frac{d-1}{d\ln 2}\bigl(\ln x(t)+1\bigr)
  +\frac{d-1}{d\ln 2}\bigl(\ln y(t)+1\bigr)\\
&=-\frac{d-1}{d\ln 2}
  \ln\!\left(\frac{x(t)}{y(t)}\right)
=-\frac{d-1}{d\ln 2}L(t).
\end{align*}
The chain rule therefore gives
\[
s_d'(P)
=F_d'(t)\frac{dt}{dP}
=-\frac{L(t)}{2t\ln 2}.
\]
For $0<t<1$,
\[
\frac{1+(d-1)t}{1-t}
=1+\frac{dt}{1-t}>1,
\]
so $L(t)>0$. Consequently,
\[
s_d'(P)<0
\qquad \text{for }1/d<P<1.
\]
For any $1/d\le P_1<P_2\le1$, continuity on
$[P_1,P_2]$ and differentiability on $(P_1,P_2)$
allow us to apply the mean value theorem:
\[
s_d(P_2)-s_d(P_1)
=s_d'(c)(P_2-P_1)<0
\]
for some $c\in(P_1,P_2)$.
Hence $s_d$ is strictly decreasing on $[1/d,1]$.

\medskip
\noindent\underline{\textbf{Change of concavity.}}
Differentiating once more yields
\[
s_d''(P)
=
\frac{d}{4(d-1)t^3\ln 2}
\bigl[L(t)-tL'(t)\bigr].
\]
Define
\[
A(t)=L(t)-tL'(t).
\]
The prefactor in the expression for $s_d''(P)$ is
positive, so $s_d''(P)$ has the same sign as $A(t)$.
Moreover,
\[
A'(t)=-tL''(t),
\qquad
L''(t)=\frac{1}{(1-t)^2}
-\frac{(d-1)^2}{[1+(d-1)t]^2}.
\]
Since $L(0)=0$ and $L'(0)=d$, the continuous
extension of $A$ satisfies $A(0)=0$.

For $d=2$, we have
\[
L''(t)=\frac{1}{(1-t)^2}-\frac{1}{(1+t)^2}>0
\qquad (0<t<1).
\]
Thus $A'(t)<0$ and $A(t)<0$ on $(0,1)$.
It follows that $s_2''(P)<0$ throughout $(1/2,1)$.

Now suppose $d>2$. For $0<t<1$, factoring the
difference of squares gives
\begin{align*}
L''(t)
&=
\left(\frac{1}{1-t}
-\frac{d-1}{1+(d-1)t}\right)
\left(\frac{1}{1-t}
+\frac{d-1}{1+(d-1)t}\right)\\
&=
\frac{2(d-1)t-(d-2)}
     {(1-t)[1+(d-1)t]}
\cdot
\frac{d}{(1-t)[1+(d-1)t]}\\
&=
\frac{d\,[2(d-1)t-(d-2)]}
     {(1-t)^2[1+(d-1)t]^2}.
\end{align*}
Since the denominator and $d$ are strictly positive,
\[
\operatorname{sgn}L''(t)
=
\operatorname{sgn}\bigl(2(d-1)t-(d-2)\bigr).
\]

Hence, with
\[
t_0=\frac{d-2}{2(d-1)},
\]
we have $A'(t)>0$ on $(0,t_0)$ and $A'(t)<0$
on $(t_0,1)$. In particular, $A(t_0)>0$.

To determine the behavior near $1$, put $r=1-t$.
Then
\[
A(1-r)
=
\ln[d-(d-1)r]-\ln r
-\frac{(d-1)(1-r)}{d-(d-1)r}
-\frac{1-r}{r}.
\]
Using $\lim_{r\to0^+}r\ln r=0$, we obtain
\begin{align*}
\lim_{r\to0^+}rA(1-r)
&=\lim_{r\to0^+}
\left[
r\ln[d-(d-1)r]-r\ln r
-\frac{r(d-1)(1-r)}{d-(d-1)r}
-(1-r)
\right]\\
&=-1.
\end{align*}
Therefore $A(t)\to-\infty$ as $t\to1^-$.

By continuity and strict decrease on $(t_0,1)$,
there is exactly one $t_*\in(t_0,1)$ such that
$A(t_*)=0$. Moreover,
\[
A(t)>0\quad(0<t<t_*),
\qquad
A(t)<0\quad(t_*<t<1).
\]
Define
\[
P_*=\frac{1}{d}+\frac{d-1}{d}t_*^2.
\]
Since $P_d(t)$ is strictly increasing, we conclude that
\[
s_d''(P)>0\quad(1/d<P<P_*),
\qquad
s_d''(P)<0\quad(P_*<P<1).
\]

\medskip
\noindent\underline{\textbf{Maximization.}}
Fix $0<\epsilon<(d-1)/d$.
After interchanging the two states if necessary,
their purities can be written as $P$ and $P+\epsilon$,
where
\[
\frac{1}{d}\le P\le1-\epsilon.
\]
Since $s_d$ is decreasing, their absolute entropy
difference equals
\[
G_\epsilon(P)=s_d(P)-s_d(P+\epsilon).
\]
At interior points of the allowed interval,
\[
G_\epsilon'(P)=s_d'(P)-s_d'(P+\epsilon),
\qquad
G_\epsilon''(P)=s_d''(P)-s_d''(P+\epsilon).
\]

For $d=2$, the strict concavity of $s_2$ implies
that $s_2'$ is strictly decreasing. Hence
$G_\epsilon'(P)>0$, so the maximum occurs at
$P=1-\epsilon$.

For $d>2$, consider the following three cases:
\begin{enumerate}
\item If $P+\epsilon\le P_*$, then $s_d'$ is
strictly increasing between the two arguments.
Thus
\[
G_\epsilon'(P)<0.
\]

\item If $P<P_*<P+\epsilon$, then
\[
G_\epsilon''(P)
=s_d''(P)-s_d''(P+\epsilon)>0.
\]
Consequently, any stationary point in this region
is a strict local minimum.

\item If $P\ge P_*$, then $s_d'$ is strictly
decreasing between the two arguments. Thus
\[
G_\epsilon'(P)>0.
\]
\end{enumerate}
These regions occur in the listed order as $P$
increases; some may be empty. If a transition
point lies in the interior of the allowed interval,
then $G_\epsilon'(P)<0$ at $P+\epsilon=P_*$
and $G_\epsilon'(P)>0$ at $P=P_*$.
Thus transition points cannot be interior maxima
either.

It follows that $G_\epsilon$ has no interior
maximum. Since it is continuous on
$[1/d,1-\epsilon]$, its maximum is attained at
an endpoint. Therefore, for every $d\ge2$,
\[
G_\epsilon(P)
\le
\max\left\{
G_\epsilon(1/d),\,G_\epsilon(1-\epsilon)
\right\}.
\]
Using
\[
a_d(\epsilon)=\sqrt{\frac{d\epsilon}{d-1}},
\qquad
b_d(\epsilon)=\sqrt{1-\frac{d\epsilon}{d-1}},
\]
the endpoint values are
\[
G_\epsilon(1/d)
=\log_2d-F_d(a_d(\epsilon)),
\qquad
G_\epsilon(1-\epsilon)
=F_d(b_d(\epsilon)).
\]
Their maximum is $B_d(\epsilon)$.
Without loss of generality, suppose that
\[
    \operatorname{Tr}(\rho_t^2)
    \geq \operatorname{Tr}(\rho_u^2).
\]
Let
\[
    P=\operatorname{Tr}(\rho_u^2),
    \qquad
    \operatorname{Tr}(\rho_t^2)=P+\epsilon.
\]
Since both purities belong to $[1/d,1]$, we have
\[
    \frac{1}{d}\leq P\leq 1-\epsilon.
\]
Moreover, because $s_d$ is decreasing,
\[
    |S(\rho_t)-S(\rho_u)|
    =|s_d(P+\epsilon)-s_d(P)|
    =s_d(P)-s_d(P+\epsilon)
    =G_\epsilon(P).
\]

As established above, $G_\epsilon$ attains its
maximum at an endpoint of $[1/d,1-\epsilon]$.
Consequently,
\begin{align*}
    |S(\rho_t)-S(\rho_u)|
    &=G_\epsilon(P)\\
    &\leq
    \max_{Q\in[1/d,\,1-\epsilon]}G_\epsilon(Q)\\
    &=\max\left\{
        G_\epsilon\left(\frac{1}{d}\right),
        G_\epsilon(1-\epsilon)
    \right\}.
\end{align*}

To evaluate these endpoints, recall that
\[
    s_d(Q)
    =F_d\left(\sqrt{\frac{dQ-1}{d-1}}\right),
    \qquad
    F_d(0)=\log_2 d,
    \qquad
    F_d(1)=0.
\]
At the left endpoint,
\begin{align*}
    G_\epsilon\left(\frac{1}{d}\right)
    &=s_d\left(\frac{1}{d}\right)
      -s_d\left(\frac{1}{d}+\epsilon\right)\\
    &=F_d(0)
      -F_d\left(
        \sqrt{\frac{d(1/d+\epsilon)-1}{d-1}}
      \right)\\
    &=\log_2 d
      -F_d\left(\sqrt{\frac{d\epsilon}{d-1}}\right)\\
    &=\log_2 d-F_d(a_d(\epsilon)).
\end{align*}
At the right endpoint,
\begin{align*}
    G_\epsilon(1-\epsilon)
    &=s_d(1-\epsilon)-s_d(1)\\
    &=F_d\left(
        \sqrt{\frac{d(1-\epsilon)-1}{d-1}}
      \right)-F_d(1)\\
    &=F_d\left(
        \sqrt{1-\frac{d\epsilon}{d-1}}
      \right)\\
    &=F_d(b_d(\epsilon)).
\end{align*}
Here,
\[
    a_d(\epsilon)=\sqrt{\frac{d\epsilon}{d-1}},
    \qquad
    b_d(\epsilon)=\sqrt{1-\frac{d\epsilon}{d-1}}.
\]
Substituting these endpoint values yields
\begin{align*}
    |S(\rho_t)-S(\rho_u)|
    &\leq
    \max\left\{
        \log_2 d-F_d(a_d(\epsilon)),
        F_d(b_d(\epsilon))
    \right\}\\
    &=B_d(\epsilon).
\end{align*}
This proves the claimed inequality for
$0<\epsilon<(d-1)/d$.
%%%%%%%%%%%%%%%%%%%%%%%

\medskip
\noindent\underline{\textbf{Endpoint cases and sharpness.}}
If $\epsilon=0$, equal purity within this family
implies equal entropy, and $B_d(0)=0$.
If $\epsilon=(d-1)/d$, the two purities must be
$1/d$ and $1$, so the entropy difference is
$\log_2d=B_d((d-1)/d)$.

For every admissible $\epsilon$, the two endpoint
values are attained, respectively, by
\[
(\rho_0,\rho_{a_d(\epsilon)})
\qquad\text{and}\qquad
(\rho_{b_d(\epsilon)},\rho_1).
\]
Indeed,
\[
P_d(a_d(\epsilon))-P_d(0)=\epsilon,
\qquad
P_d(1)-P_d(b_d(\epsilon))=\epsilon.
\]
Choosing the pair corresponding to the larger
endpoint value establishes sharpness.
\end{proof}

For $d=2$, the second endpoint gives the maximum,
and the bound reduces to
\[
B_2(\epsilon)
=
h\!\left(\frac{1-\sqrt{1-2\epsilon}}{2}\right),
\qquad 0\leq\epsilon\leq\frac12,
\]
where $h$ denotes the binary entropy.
For $d>2$, the entropy--purity relation changes
from convex to concave, and the sharp bound is
obtained by taking the larger of the two endpoint
values in Eq.~\eqref{eq:Bd-definition}.

We now apply this result to bipartite systems with
$\dim A=\dim B=m\geq2$, so that the total dimension
is $d=m^2$. Let
\[
|\Phi_m\rangle
=
\frac{1}{\sqrt m}\sum_{j=1}^{m}|jj\rangle,
\]
and consider the one-parameter family of isotropic
states
\begin{equation}
\omega_t^{AB}
=
t|\Phi_m\rangle\langle\Phi_m|
+
(1-t)\frac{I_{AB}}{m^2},
\qquad 0\leq t\leq1.
\label{eq:isotropic-family}
\end{equation}
This is a special case of the preceding family
with $d=m^2$. Since
\[
\operatorname{Tr}_B
\bigl(|\Phi_m\rangle\langle\Phi_m|\bigr)
=\frac{I_A}{m},
\qquad
\operatorname{Tr}_B
\left(\frac{I_{AB}}{m^2}\right)
=\frac{I_A}{m},
\]
and similarly for the partial trace over $A$,
both marginals are maximally mixed:
\[
\omega_t^A=\frac{I_A}{m},
\qquad
\omega_t^B=\frac{I_B}{m}.
\]

We now define the entropic quantities considered below.
Let $\omega^{AB}$ be a density operator on
$\mathcal{H}_A\otimes\mathcal{H}_B$, with reduced states
$\omega^A=\operatorname{Tr}_B\omega^{AB}$ and
$\omega^B=\operatorname{Tr}_A\omega^{AB}$.
All logarithms are taken to base $2$.

\begin{itemize}
    \item \textbf{Von Neumann entropy.}
    The joint entropy of $\omega^{AB}$ is
    \[
    S(AB)_\omega
    :=-\operatorname{Tr}\!\left(
    \omega^{AB}\log_2\omega^{AB}\right),
    \]
    where $0\log_2 0:=0$.

    \item \textbf{Conditional von Neumann entropy.}
    The conditional entropy of $A$ given $B$ is
    \[
    S(A|B)_\omega
    :=S(AB)_\omega-S(B)_\omega.
    \]

    \item \textbf{Quantum mutual information.}
    The mutual information between $A$ and $B$ is
    \[
    I(A:B)_\omega
    :=S(A)_\omega+S(B)_\omega-S(AB)_\omega.
    \]

    \item \textbf{Coherent information.}
    The coherent information from $A$ to $B$ is
    \[
    I(A\rangle B)_\omega
    :=S(B)_\omega-S(AB)_\omega
    =-S(A|B)_\omega.
    \]

    \item \textbf{Relative Entropy Distances.}
    We define
    \[
    D_{\mathcal E}(\omega)
    :=\inf_{\tau_B\in\mathcal{D}(\mathcal{H}_B)}
    D\!\left(\omega^{AB}\middle\|I_A\otimes\tau_B\right),
    \]
    where $\mathcal E=\{I_A\otimes\tau_B:\tau_B\in\mathrm{D}(B)\}$ and  $\mathcal{D}(\mathcal{H}_B)$ denotes the set of density operators on $\mathcal{H}_B$. Here \[D(\rho\|\sigma)=\operatorname{Tr}\bigl[\rho(\log_2\rho-\log_2\sigma)\bigr]
    \] when $\operatorname{supp}\rho\subseteq\operatorname{supp}\sigma$ and $D(\rho\|\sigma)=+\infty$ otherwise. In the definition of $D_{\mathcal E}$, the second argument is unnormalized.

    \item \textbf{Quantum conditional mutual information.}
    For a tripartite state $\omega^{ABC}$, the quantum
    conditional mutual information between $A$ and $B$
    given $C$ is defined as
    \[
    I(A:B|C)_\omega
    :=S(AC)_\omega+S(BC)_\omega
    -S(C)_\omega-S(ABC)_\omega.
    \]
    Equivalently,
    \[
    I(A:B|C)_\omega
    =S(A|C)_\omega-S(A|BC)_\omega.
    \]
    By strong subadditivity of the von Neumann entropy,
    $I(A:B|C)_\omega\geq 0$.

    \item \textbf{Squashed entanglement.}
    For a bipartite state $\omega^{AB}$, the squashed
    entanglement is defined as
    \[
    E_{\mathrm{sq}}(A:B)_\omega
    :=\frac{1}{2}
    \inf_{\substack{\omega^{ABC}\\
    \operatorname{Tr}_C\omega^{ABC}=\omega^{AB}}}
    I(A:B|C)_\omega,
    \]
    where the infimum is taken over all extensions
    $\omega^{ABC}$ of $\omega^{AB}$, allowing arbitrary
    auxiliary systems $C$, and $I(A:B|C)_\omega$ denotes
    the quantum conditional mutual information.

    % \item \textbf{Quantum Markov chain.}
    % A tripartite state $\omega^{ABC}$ is called a quantum
    % Markov chain in the order $A\leftrightarrow C \leftrightarrow B$ if
    % \[
    % I(A:B|C)_\omega=0.
    % \]
    % Equivalently, there exists a completely positive and
    % trace-preserving recovery map
    % \[
    % \mathcal{R}_{C\to BC}:
    % \mathcal{L}(\mathcal{H}_C)
    % \longrightarrow
    % \mathcal{L}(\mathcal{H}_B\otimes\mathcal{H}_C)
    % \]
    % such that
    % \[
    % \omega^{ABC}
    % =
    % \bigl(\operatorname{id}_A\otimes
    % \mathcal{R}_{C\to BC}\bigr)(\omega^{AC}),
    % \]
    % where $\omega^{AC}=\operatorname{Tr}_B\omega^{ABC}$,
    % $\operatorname{id}_A$ is the identity channel on $A$,
    % and $\mathcal{L}(\mathcal{H})$ denotes the space of
    % linear operators on $\mathcal{H}$.
    % Thus, $\omega^{ABC}$ can be recovered from $\omega^{AC}$
    % by acting only on the conditioning system $C$.
    
\end{itemize}

\begin{theorem}[Conditional entropy, mutual information,
and relative-entropy functionals]
\label{thm:isotropic-quantities}
Let $\omega_t,\omega_u$ belong to
Eq.~\eqref{eq:isotropic-family}, and
$\epsilon
=
\left|
\operatorname{Tr}(\omega_t^2)
-
\operatorname{Tr}(\omega_u^2)
\right|.$
For each functional
\[
Q\in
\left\{
S(AB),\,
S(A|B),\,
I(A:B),\,
I(A\rangle B),\,
D_{\mathcal E}
\right\},
\]
one has the sharp bound
\begin{equation}
|Q(\omega_t)-Q(\omega_u)|
\leq B_{m^2}(\epsilon).
\label{eq:isotropic-all-bounds}
\end{equation}

\end{theorem}

\begin{proof}
The spectrum of $\omega_t^{AB}$ gives
$S(AB)_{\omega_t}=F_{m^2}(t)$, while its maximally mixed
marginals satisfy
\[
S(A)_{\omega_t}=S(B)_{\omega_t}=\log_2 m.
\]
Substituting these identities into the definitions yields
\begin{align*}
S(AB)_{\omega_t}
&=F_{m^2}(t),\\
S(A|B)_{\omega_t}
&=F_{m^2}(t)-\log_2 m,\\
I(A:B)_{\omega_t}
&=2\log_2 m-F_{m^2}(t),\\
I(A\rangle B)_{\omega_t}
&=\log_2 m-F_{m^2}(t).
\end{align*}

Moreover, for any bipartite state $\omega^{AB}$,
\[
D(\omega^{AB}\|I_A\otimes\tau_B)
=
-S(A|B)_\omega+D(\omega_B\|\tau_B).
\]
Since $D(\omega_B\|\tau_B)\geq0$, with equality at
$\tau_B=\omega_B$, minimization over density operators
$\tau_B$ gives
\[
D_{\mathcal E}(\omega)=-S(A|B)_\omega.
\]
In particular,
\[
D_{\mathcal E}(\omega_t)
=
\log_2 m-F_{m^2}(t).
\]

Thus each quantity $Q$ considered above has the form
$Q(\omega_t)=c_Q\pm F_{m^2}(t)$, where $c_Q$ is independent
of $t$. Consequently,
\[
|Q(\omega_t)-Q(\omega_u)|
=
|F_{m^2}(t)-F_{m^2}(u)|.
\]
Applying Theorem~\ref{thm:family-entropy} with $d=m^2$
proves all the bounds in
Eq.~\eqref{eq:isotropic-all-bounds}.
The same pairs of states that attain the entropy bound
also attain each of these bounds, establishing sharpness.
\end{proof}

The elements of $\mathcal E$ above are positive
operators of trace $m$, rather than normalized
states. Accordingly, $D_{\mathcal E}$ need not
be nonnegative. The preceding theorem concerns
this specified set $\mathcal E$, not an arbitrary
relative-entropy distance.

For a tripartite application, let
$\dim A=\dim B=\dim C=m$, and define
\begin{equation}
\gamma_t^{ABC}
=
t|\mathrm{GHZ}_m\rangle\langle\mathrm{GHZ}_m|
+
(1-t)\frac{I_{ABC}}{m^3},
\qquad
|\mathrm{GHZ}_m\rangle
=
\frac1{\sqrt m}\sum_{j=1}^{m}|jjj\rangle.
\label{eq:GHZ-family}
\end{equation}

\begin{theorem}[Conditional mutual information]
\label{thm:ghz-cmi}
Let $m\geq2$, and let $\gamma_t^{ABC}$ and
$\gamma_u^{ABC}$ be the states defined in
Eq.~\eqref{eq:GHZ-family}, with $t,u\in[0,1]$.
Then
\begin{equation}
I(A:B|C)_{\gamma_t}
=
\log_2 m+2F_m(t)-F_{m^3}(t).
\label{eq:ghz-cmi-formula}
\end{equation}
Define
\[
\epsilon
:=
\left|
\operatorname{Tr}\bigl[(\gamma_t^{ABC})^2\bigr]
-
\operatorname{Tr}\bigl[(\gamma_u^{ABC})^2\bigr]
\right|,
\qquad
c_m:=\frac{m^2}{m^2+m+1}.
\]
Then
\begin{equation}
\left|
I(A:B|C)_{\gamma_t}
-
I(A:B|C)_{\gamma_u}
\right|
\leq
\min\left\{
2\log_2 m,\,
2B_m(c_m\epsilon)+B_{m^3}(\epsilon)
\right\}.
\label{eq:ghz-cmi-bound}
\end{equation}
\end{theorem}

\begin{proof}
Tracing out $AB$ gives $\gamma_t^C=I_C/m$, and hence
\[
S(C)_{\gamma_t}=\log_2 m.
\]
Moreover,
\[
\gamma_t^{AC}
=
\frac{t}{m}\sum_{j=1}^{m}|jj\rangle\langle jj|
+
\frac{1-t}{m^2}I_{AC}.
\]
Its eigenvalues, counted with multiplicity, are
\[
\frac{1+(m-1)t}{m^2}
\quad\text{with multiplicity }m,
\qquad
\frac{1-t}{m^2}
\quad\text{with multiplicity }m(m-1).
\]
These are the eigenvalues defining $F_m(t)$, each
divided by $m$ and repeated $m$ times. Consequently,
\[
S(AC)_{\gamma_t}=\log_2 m+F_m(t).
\]
By symmetry, the same identity holds for
$S(BC)_{\gamma_t}$. Since
$S(ABC)_{\gamma_t}=F_{m^3}(t)$, we obtain
\begin{align*}
I(A:B|C)_{\gamma_t}
&=S(AC)_{\gamma_t}+S(BC)_{\gamma_t}
  -S(C)_{\gamma_t}-S(ABC)_{\gamma_t}\\
&=\log_2 m+2F_m(t)-F_{m^3}(t),
\end{align*}

Recall that the purity of the $d$-dimensional family is
\[
P_d(t)=\frac{1}{d}+\frac{d-1}{d}t^2.
\]
Therefore,
\[
\epsilon
=
|P_{m^3}(t)-P_{m^3}(u)|
=
\frac{m^3-1}{m^3}|t^2-u^2|,
\]
while
\[
|P_m(t)-P_m(u)|
=
\frac{m-1}{m}|t^2-u^2|
=
c_m\epsilon.
\]
In particular,
\[
0\leq\epsilon\leq1-\frac{1}{m^3},
\qquad
0\leq c_m\epsilon\leq1-\frac{1}{m},
\]
so both arguments lie in the domains of the
corresponding entropy bounds.

Applying Theorem~\ref{thm:family-entropy} in dimensions
$m$ and $m^3$, respectively, gives
\[
|F_m(t)-F_m(u)|\leq B_m(c_m\epsilon),
\qquad
|F_{m^3}(t)-F_{m^3}(u)|
\leq B_{m^3}(\epsilon).
\]
Using Eq.~\eqref{eq:ghz-cmi-formula} and the triangle
inequality, we conclude that
\begin{align*}
\left|
I(A:B|C)_{\gamma_t}
-
I(A:B|C)_{\gamma_u}
\right|
&\leq
2|F_m(t)-F_m(u)|
+
|F_{m^3}(t)-F_{m^3}(u)|\\
&\leq
2B_m(c_m\epsilon)+B_{m^3}(\epsilon).
\end{align*}

Finally, for any tripartite state $\omega^{ABC}$ with
$\dim A=m$, strong subadditivity and the conditional
entropy bounds
\[
-\log_2 m\leq S(A|X)_\omega\leq\log_2 m,
\qquad X=C,BC,
\]
imply
\[
0\leq I(A:B|C)_\omega
=S(A|C)_\omega-S(A|BC)_\omega
\leq2\log_2 m.
\]
Thus,
\[
\left|
I(A:B|C)_{\gamma_t}
-
I(A:B|C)_{\gamma_u}
\right|
\leq2\log_2 m.
\]
Combining the two inequalies proves
Eq.~\eqref{eq:ghz-cmi-bound}.
\end{proof}

% \begin{theorem}[Comparison with a Markov state]
% \label{thm:GHZ-Markov}
% The state $\gamma_0=I_{ABC}/m^3$ is a quantum
% Markov state. For
% \[
% \epsilon=\operatorname{Tr}(\gamma_t^2)-\frac1{m^3},
% \qquad
% a=\sqrt{\frac{m^3\epsilon}{m^3-1}},
% \]
% one has
% \begin{equation}
% I(A:B|C)_{\gamma_t}
% =
% \log_2m+2F_m(a)-F_{m^3}(a).
% \label{eq:Markov-exact}
% \end{equation}
% In particular, this conditional mutual information
% converges to zero as $\epsilon\to0$.
% \end{theorem}

% \begin{proof}
% Since
% \[
% \gamma_0
% =
% \frac{I_A}{m}\otimes
% \frac{I_B}{m}\otimes
% \frac{I_C}{m},
% \]
% we have
% \[
% I(A:B|C)_{\gamma_0}
% =
% 2\log_2m+2\log_2m-\log_2m-3\log_2m
% =
% 0.
% \]
% Hence $\gamma_0$ is a quantum Markov state
% in the order $A-C-B$.

% The purity formula gives
% \[
% \epsilon
% =
% \operatorname{Tr}(\gamma_t^2)-\frac{1}{m^3}
% =
% \frac{m^3-1}{m^3}t^2.
% \]
% Since $t\in[0,1]$, it follows that
% \[
% t=\sqrt{\frac{m^3\epsilon}{m^3-1}}=a.
% \]
% Substituting $t=a$ into
% Eq.~\eqref{eq:ghz-cmi-formula} yields
% \[
% I(A:B|C)_{\gamma_t}
% =
% \log_2m+2F_m(a)-F_{m^3}(a),
% \]
% which proves Eq.~\eqref{eq:Markov-exact}.

% Finally, as $\epsilon\to0^+$, we have $a\to0$.
% Since $F_d$ is continuous at zero and
% $F_d(0)=\log_2d$, we obtain
% \[
% \begin{aligned}
% \lim_{\epsilon\to0^+}I(A:B|C)_{\gamma_t}
% &=
% \log_2m+2F_m(0)-F_{m^3}(0)\\
% &=
% \log_2m+2\log_2m-\log_2(m^3)\\
% &=0.
% \end{aligned}
% \]
% \end{proof}

Finally, consider squashed entanglement,
\[
E_{\mathrm{sq}}(A:B)_\omega
=
\frac12
\inf_{\omega^{ABE}:\operatorname{Tr}_E\omega^{ABE}
=\omega^{AB}}
I(A:B|E)_\omega.
\]

\begin{theorem}[Squashed-entanglement continuity bound]
\label{thm:isotropic-squashed}
Let $m\geq2$, and let $\omega_t^{AB}$ and
$\omega_u^{AB}$ belong to the fixed isotropic family
in Eq.~\eqref{eq:isotropic-family}, with $t,u\in[0,1]$.
Let $D=m^2$ and define
\[
\epsilon
:=
\left|
\operatorname{Tr}(\omega_t^2)
-
\operatorname{Tr}(\omega_u^2)
\right|,
\qquad
x:=\sqrt{\frac{D-1}{D}\epsilon},
\qquad
z:=\min\{1,\sqrt{2x}\}.
\]
Then
\begin{equation}
\begin{split}
&\left|
E_{\mathrm{sq}}(\omega_t)
-
E_{\mathrm{sq}}(\omega_u)
\right|\\
&\quad\leq
\min\left\{
\log_2 m,\,
2z\log_2 m+
(1+z)h\left(\frac{z}{1+z}\right)
\right\},
\end{split}
\label{eq:squashed-family-bound}
\end{equation}
where
$h(p)=-p\log_2p-(1-p)\log_2(1-p)$
is the binary entropy, with $0\log_2 0:=0$.
In particular, the right-hand side converges to zero
as $\epsilon\to0$.
\end{theorem}

\begin{proof}
Let $\Pi=|\Phi_m\rangle\langle\Phi_m|$, so that
\[
\omega_t=t\Pi+(1-t)\frac{I_{AB}}{D}.
\]
Using $\Pi^2=\Pi$ and $\operatorname{Tr}\Pi=1$,
we obtain
\[
\operatorname{Tr}(\omega_t^2)
=
t^2+\frac{2t(1-t)}{D}+\frac{(1-t)^2}{D}
=
\frac1D+\frac{D-1}{D}t^2.
\]
Consequently,
\[
\epsilon=\frac{D-1}{D}|t^2-u^2|.
\]

Now,
\[
\omega_t-\omega_u
=(t-u)\left(\Pi-\frac{I_{AB}}{D}\right).
\]
The operator $\Pi-I_{AB}/D$ has eigenvalues
$(D-1)/D$ and $-1/D$ with multiplicity $D-1$.
Thus,
\[
T:=\frac12\|\omega_t-\omega_u\|_1
=\frac{D-1}{D}|t-u|.
\]
As $t,u\geq0$, we have
\[
|t-u|^2
\leq |t-u|(t+u)
=|t^2-u^2|.
\]
It follows that
\[
T^2
\leq
\left(\frac{D-1}{D}\right)^2|t^2-u^2|
=
\frac{D-1}{D}\epsilon=x^2,
\]
and hence $T\leq x$.

Let
\[
f(\rho,\sigma):=\|\sqrt{\rho}\sqrt{\sigma}\|_1
\]
denote the root fidelity.
By Uhlmann's theorem \cite{Uhlmann1976}, purifications
$\psi_t^{ABR}$ and $\psi_u^{ABR}$ can be chosen so that
\[
\frac12\|\psi_t-\psi_u\|_1
=
\sqrt{1-f(\omega_t,\omega_u)^2}.
\]
The fidelity--trace-distance inequality
$f(\omega_t,\omega_u)\geq1-T$ implies
\[
\frac12\|\psi_t-\psi_u\|_1
\leq\sqrt{2T-T^2}
\leq\sqrt{2T}.
\]
Since trace distance is at most one, we conclude that
\[
\frac12\|\psi_t-\psi_u\|_1
\leq\min\{1,\sqrt{2T}\}\leq z.
\]

Every extension $\eta^{ABE}$ of $\omega_t^{AB}$
can be obtained by applying a channel
$\Lambda_{R\to E}$ to its purifying system:
\[
\eta^{ABE}
=
(\operatorname{id}_{AB}\otimes\Lambda_{R\to E})
(\psi_t^{ABR}).
\]
Applying the same channel to $\psi_u^{ABR}$ gives
an extension $\xi^{ABE}$ of $\omega_u^{AB}$.
Contractivity of trace distance yields
\[
\frac12\|\eta-\xi\|_1\leq z.
\]

Define
\[
K_m(z)
:=
2z\log_2m+
(1+z)h\left(\frac{z}{1+z}\right).
\]
Winter's conditional-entropy continuity bound
\cite{Winter2016}, together with contractivity under
partial trace, gives
\[
|S(A|E)_\eta-S(A|E)_\xi|\leq K_m(z),
\qquad
|S(A|BE)_\eta-S(A|BE)_\xi|\leq K_m(z).
\]
Using
\[
I(A:B|E)=S(A|E)-S(A|BE),
\]
we obtain
\[
\frac12
\left|I(A:B|E)_\eta-I(A:B|E)_\xi\right|
\leq K_m(z).
\]

For any $\delta>0$, choose an extension $\eta^{ABE}$
of $\omega_t$ satisfying
\[
\frac12 I(A:B|E)_\eta
\leq E_{\mathrm{sq}}(\omega_t)+\delta.
\]
For the corresponding extension $\xi^{ABE}$
of $\omega_u$, we have
\begin{align*}
E_{\mathrm{sq}}(\omega_u)
&\leq \frac12 I(A:B|E)_\xi\\
&\leq \frac12 I(A:B|E)_\eta+K_m(z)\\
&\leq E_{\mathrm{sq}}(\omega_t)+\delta+K_m(z).
\end{align*}
Letting $\delta\to0$ and interchanging $t$ and $u$
therefore gives
\[
\left|
E_{\mathrm{sq}}(\omega_t)
-
E_{\mathrm{sq}}(\omega_u)
\right|
\leq K_m(z).
\]

Finally, nonnegativity of conditional mutual
information and the choice of a trivial extension give
\[
0\leq E_{\mathrm{sq}}(\omega)
\leq\frac12 I(A:B)_\omega
\leq\log_2m.
\]
Hence the absolute difference is also at most
$\log_2m$. Combining the two inequalities proves
Eq.~\eqref{eq:squashed-family-bound}.

As $\epsilon\to0$, we have $x\to0$ and $z\to0$.
Continuity of the binary entropy then shows that
the stated bound converges to zero.
\end{proof}

\section{Application to Machine-Learning-Assisted
Qubit Entropy Estimation}
\label{sec:ann-application}

\begin{figure}[!htb]
    \centering
    \includegraphics[width=\linewidth]
    {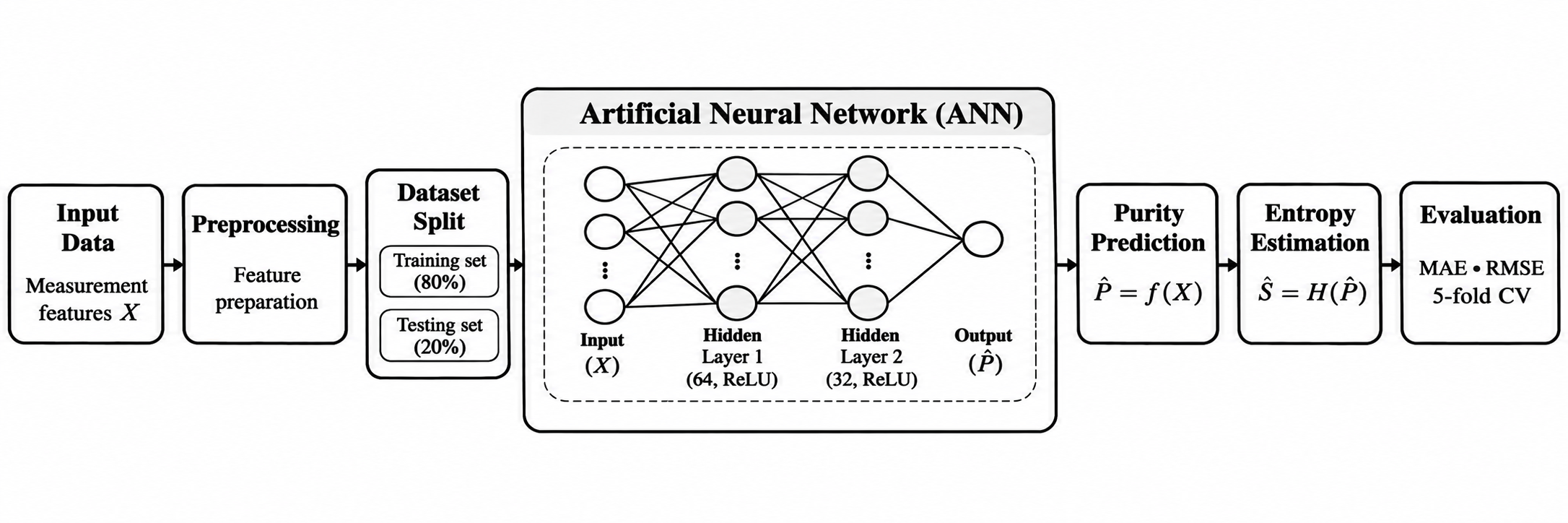}
    \caption{
    Framework for machine-learning-assisted
    qubit purity prediction and entropy
    estimation from finite Pauli measurement
    data.
    }
    \label{fig123fghhhf7894}
\end{figure}

We illustrate how the sharp qubit continuity bound
converts purity-estimation errors into rigorous
bounds on entropy-estimation errors as shown in Figure \ref{fig123fghhhf7894}. Four regression
models are considered: multilayer perceptron
(MLP or ANN), Random Forest, Extra Trees, and
XGBoost
\cite{rumelhart1986learning,breiman2001random,
geurts2006extremely,chen2016xgboost}.
Each model predicts purity from finite Pauli
measurement data, and the prediction is converted
into an entropy estimate.
For a qubit state
$\rho=(I+\mathbf r\cdot\boldsymbol{\sigma})/2$,
with $\|\mathbf r\|_2\leq1$, the purity and entropy
are
\begin{equation}
    P=\operatorname{Tr}(\rho^2)
    =\frac{1+\|\mathbf r\|_2^2}{2},
    \qquad
    S(\rho)=s(P)
    :=h\left(\frac{1-\sqrt{2P-1}}{2}\right).
    \label{eq:ann-entropy-map}
\end{equation}
For each Pauli observable $\sigma_j$,
$j\in\{x,y,z\}$, measurements are performed on
a separate batch of $N$ independently prepared
copies. The number of $+1$ outcomes and the
empirical measurement average satisfy
\begin{equation}
    K_j\sim\operatorname{Binomial}
    \left(N,\frac{1+r_j}{2}\right),
    \qquad
    X_j=\frac{2K_j}{N}-1.
    \label{eq:ann-input-average}
\end{equation}
The counts are independent conditional on the
state. The input features are
$\mathbf X=(X_x,X_y,X_z)^T$, while the exact
purity is used only as the supervised target.

States are sampled uniformly in volume from
the Bloch ball. We use $N=100$ copies per Pauli
axis, giving $300$ copies per state, with one
measurement realization per sampled state.
The development and independent test datasets
contain $15000$ and $5000$ states, generated
with seeds $2026$ and $2027$, respectively.
For the MLP fit to the development dataset,
$12750$ states are used for training and $2250$
for validation. The network has architecture
$3$--$64$--$32$--$1$, ReLU hidden activations,
and a linear output. It is trained using
squared-error loss, Adam, and validation-based
early stopping
\cite{nair2010rectified,kingma2014adam}.
Test states are excluded from training and
validation.

As illustrated in Fig.~\ref{fig123fghhhf7894},
each model prediction is restricted to the
physical purity interval before conversion
to entropy:
\begin{equation}
    \widehat P
    =\min\{1,\max\{1/2,f(\mathbf X)\}\},
    \qquad
    \widehat S=s(\widehat P).
    \label{eq:ann-estimated-entropy}
\end{equation}
Since $P\in[1/2,1]$, this clipping cannot
increase the absolute purity-estimation error.
All models are evaluated on the same test
states and measurement realizations.

For an individual state, define
\[
    e_P=|P-\widehat P|,
    \qquad
    e_S=|S(\rho)-\widehat S|.
\]
Every $\widehat P\in[1/2,1]$ corresponds to
a qubit state with entropy $s(\widehat P)$.
Therefore, Theorem~\ref{thm:sharp-qubit-bound}
and Proposition~\ref{prop:comparison-rehman}
imply
\begin{equation}
    e_S\leq B(e_P)\leq R(e_P),
    \label{eq:ann-deterministic-bound}
\end{equation}
where
\[
    B(e)=h\left(\frac{1-\sqrt{1-2e}}{2}\right),
    \qquad
    R(e)=h\left(\sqrt{\frac e2}\right),
    \qquad 0\leq e\leq\frac12.
\]
Here $R$ denotes the bound obtained in ~\cite{RehmanShin2019}. The sharp bound
$B(e)$ is strictly smaller than $R(e)$ for
$0<e<1/2$, independently of the regression model.

\begin{figure}[tb]
    \centering
    \includegraphics[width=\linewidth]
    {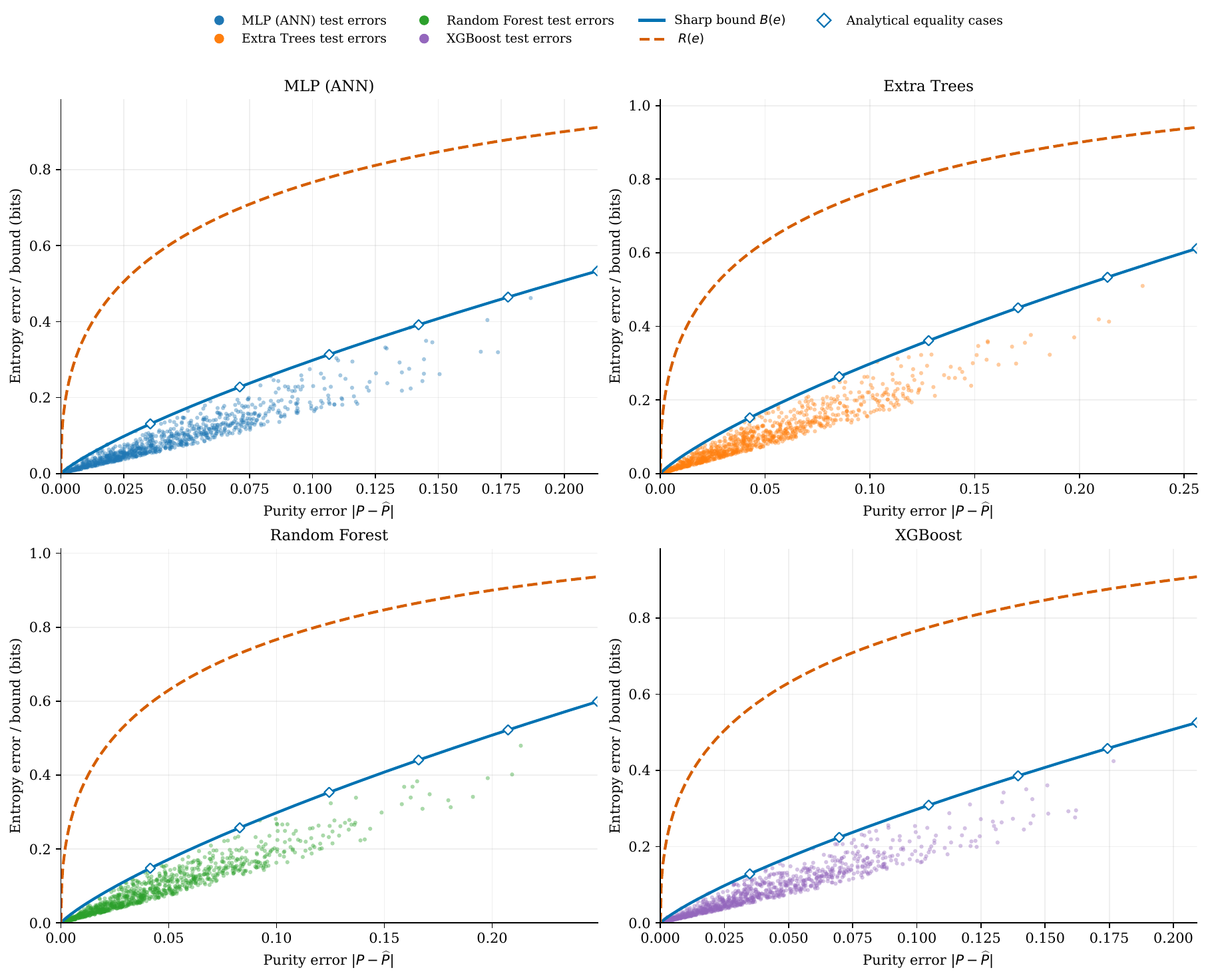}
    \caption{
    Empirical entropy errors and theoretical
    bounds for four regression models:
    (a) MLP (ANN), (b) Extra Trees,
    (c) Random Forest, and (d) XGBoost.
    Scatter points represent test-set entropy
    errors versus absolute purity-estimation
    errors. The solid blue curve denotes
    the sharp bound $B(e)$, and the dashed
    orange curve denotes the bound $R(e)$ \cite{RehmanShin2019}. Diamond markers indicate
    analytical equality cases of the sharp
    bound, rather than model predictions.
    The measurement budget is $N=100$
    copies per Pauli axis.
    }
    \label{fiG12345YU}
\end{figure}

Figure~\ref{fiG12345YU} compares the empirical
entropy errors with the theoretical bounds.
The analytical equality cases correspond to
$P=1$ and $\widehat P=1-e$, for which
$e_S=s(1-e)=B(e)$. These cases demonstrate
attainability of the sharp bound independently
of the trained models.

Predictive performance is assessed using mean
absolute error (MAE) and root mean squared
error (RMSE). For $Y\in\{P,S\}$ and an evaluation
set containing $M$ states, define
\[
    \operatorname{MAE}_Y
    =\frac1M\sum_{i=1}^{M}|Y_i-\widehat Y_i|,
    \qquad
    \operatorname{RMSE}_Y
    =\left[
        \frac1M\sum_{i=1}^{M}
        (Y_i-\widehat Y_i)^2
    \right]^{1/2}.
\]
Here $S_i=s(P_i)$ and
$\widehat S_i=s(\widehat P_i)$, and entropy
errors are measured in bits.

\begin{figure}[!htb]
    \centering
    \includegraphics[width=0.85\linewidth]
    {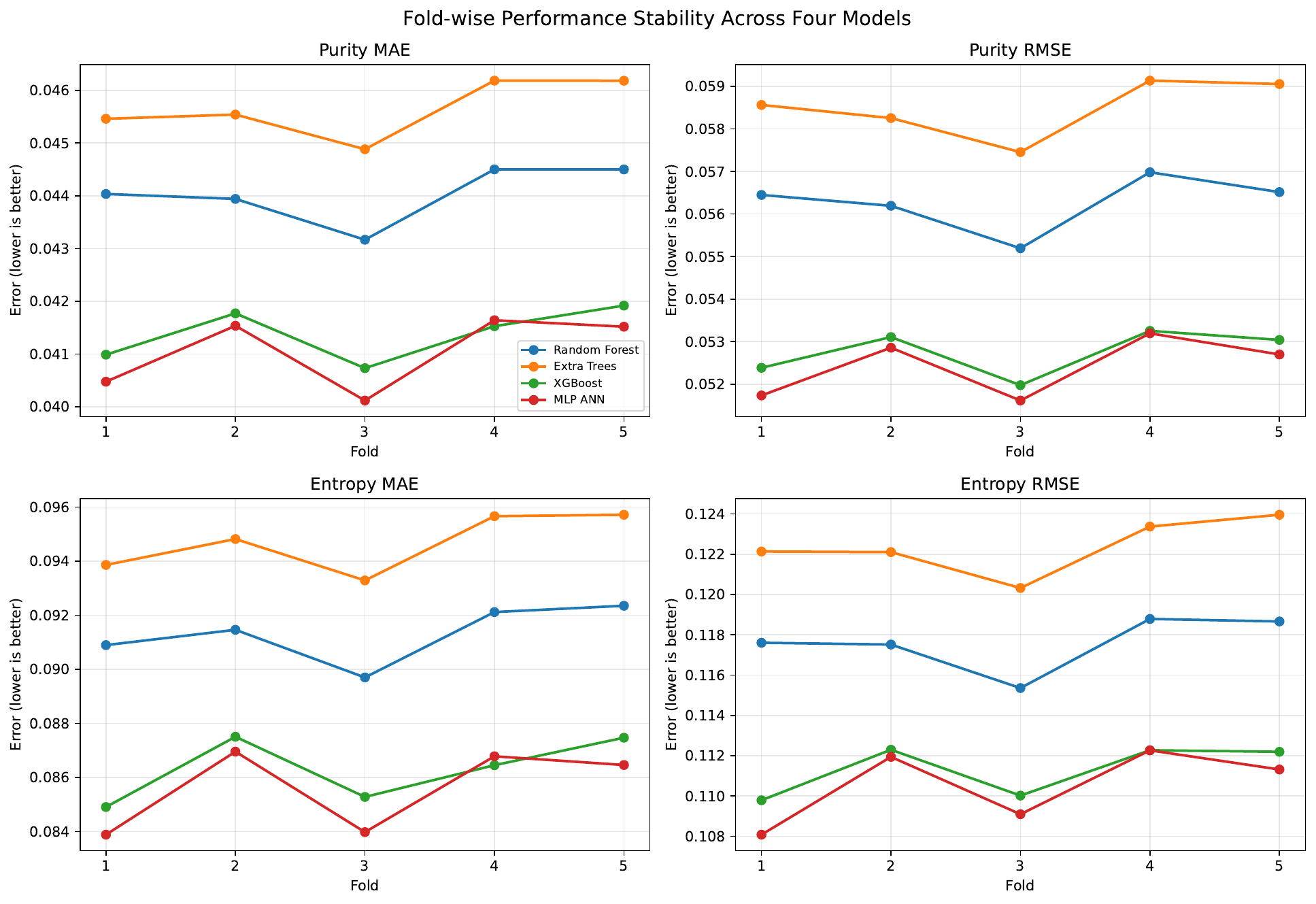}
    \caption{
    Five-fold cross-validation performance
    of Random Forest, Extra Trees, XGBoost,
    and MLP ANN for purity and entropy
    estimation. The panels show purity MAE,
    purity RMSE, entropy MAE, and entropy
    RMSE for each held-out fold.
    }
    \label{fig1234fffggfg6g56}
\end{figure}

\begin{figure}[!htb]
    \centering
    \includegraphics[width=0.85\linewidth]
    {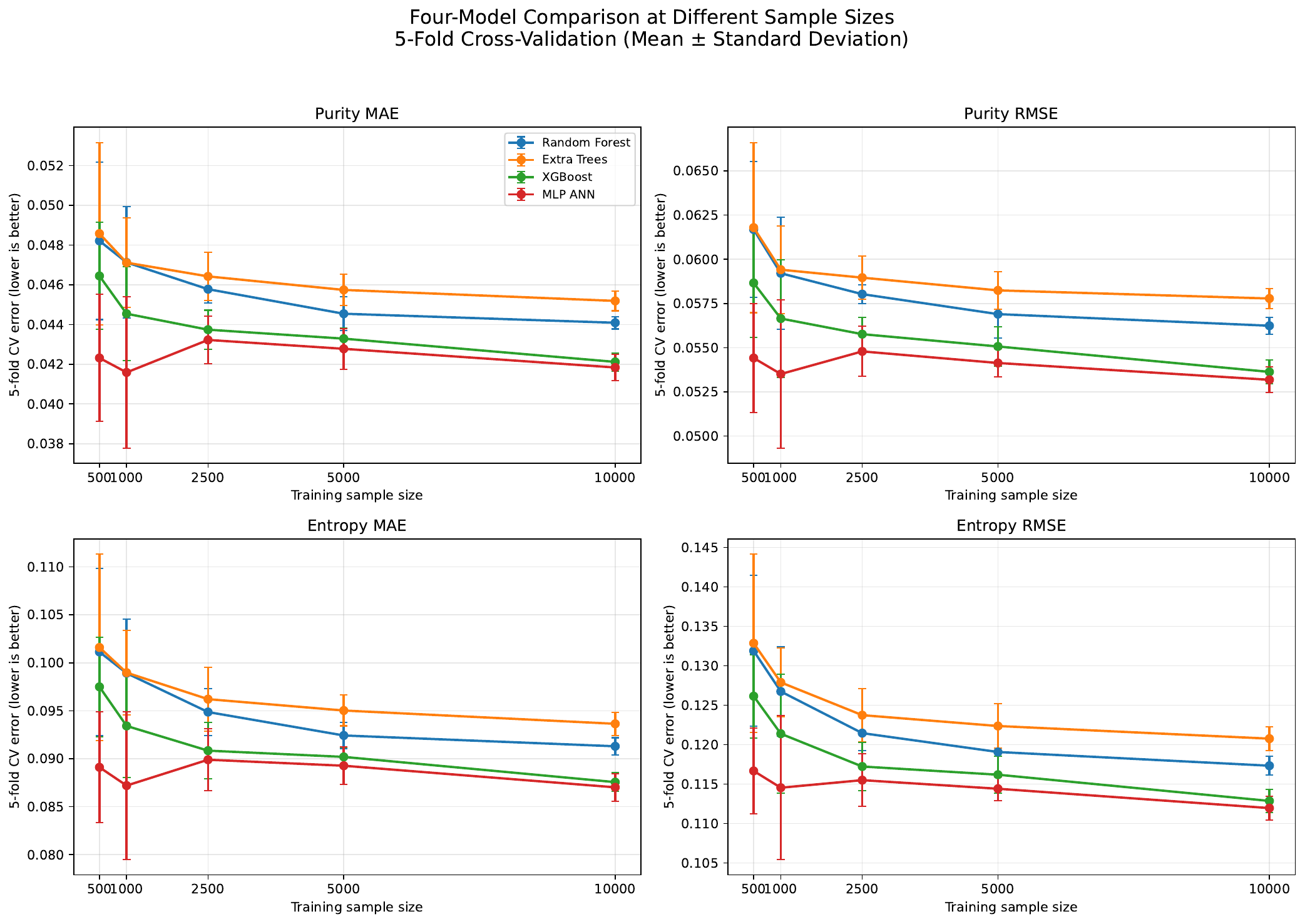}
    \caption{
    Purity- and entropy-estimation performance
    at different training sample sizes using
    five-fold cross-validation. Error bars
    indicate variability across folds.
    Lower MAE and RMSE values indicate
    better predictive performance.
    }
    \label{fig1234fghffg56}
\end{figure}

We additionally assess performance through
five-fold cross-validation
\cite{kohavi1995study,arlot2010survey}.
Each fold is held out once for evaluation,
while the remaining four folds are used for
model training.
Figure~\ref{fig1234fffggfg6g56} presents the
fold-wise errors, and
Table~\ref{456table2dddd} summarizes their means
and standard deviations.
Figure~\ref{fig1234fghffg56} examines the dependence
of predictive performance on training sample size.
The reported errors generally decrease as the
sample size increases. MLP and XGBoost perform
similarly and generally yield smaller errors
than the two other models. The error bars
describe variability across cross-validation
folds.

\begin{table}[htbp]
    \centering
    \caption{
    Sharp bound $B(e)$ purity and entropy estimation errors
using five-fold cross-validation.
    Values are means $\pm$ standard
    deviations across folds. Entropy
    errors are measured in bits.
    }
    \label{456table2dddd}
    \resizebox{\linewidth}{!}{%
    \begin{tabular}{|l|c|c|c|c|}
        \hline
        Model
        & Purity MAE
        & Purity RMSE
        & Entropy MAE
        & Entropy RMSE\\
        \hline
        Random Forest
        & $0.044028 \pm 0.000547$
        & $0.056267 \pm 0.000663$
        & $0.091307 \pm 0.001063$
        & $0.117586 \pm 0.001376$\\
        \hline
        Extra Trees
        & $0.045649 \pm 0.000549$
        & $0.058493 \pm 0.000683$
        & $0.094671 \pm 0.001081$
        & $0.122378 \pm 0.001400$\\
        \hline
        XGBoost
        & $0.041387 \pm 0.000510$
        & $0.052753 \pm 0.000547$
        & $0.086325 \pm 0.001207$
        & $0.111314 \pm 0.001290$\\
        \hline
        MLP ANN
        & $0.041057 \pm 0.000708$
        & $0.052421 \pm 0.000705$
        & $0.085614 \pm 0.001545$
        & $0.110540 \pm 0.001849$\\
        \hline
    \end{tabular}%
    }
\end{table}

The MLP achieves the lowest mean errors across
all four metrics in Table~\ref{456table2dddd},
closely followed by XGBoost. Both models have
smaller reported mean errors than Random Forest
and Extra Trees. These comparisons describe
the present numerical setting and do not
establish a statistically significant difference
between MLP and XGBoost.

The regression models supply purity estimates,
while the continuity bound controls the
corresponding entropy errors. In these simulations,
the known true purity permits direct evaluation
of $B(e_P)$. For an unknown experimental state,
a certified purity-error bound
$e_P\leq\delta\leq1/2$ implies
$e_S\leq B(\delta)$ by monotonicity of $B$.
The model prediction alone does not provide
such a certificate.

\section{Conclusion}
\label{conclusion}

We established a purity-based continuity bound for the von Neumann entropy of qubit states, attainable for every admissible purity difference. It is  strictly tighter than the existing bound. In higher dimensions, we derived a sharp bound within the family of pure states mixed with the maximally mixed state. For bipartite isotropic states, we obtained sharp bounds for conditional entropy, mutual information, coherent information, and the relative-entropy functional, together with a continuity bound for squashed entanglement. We also derived a purity-based bound for quantum conditional mutual information within the noisy generalized GHZ family.

An application to machine-learning-assisted qubit entropy
estimation demonstrated how the sharp bound converts purity-estimation errors into rigorous entropy-error bounds. Future work may investigate extensions to
general higher-dimensional states and general bipartite states, identifying the additional state information needed beyond purity to control entropy variations.

\vspace{0.5cm}
\noindent\textbf{Acknowledgements:} We thank Dr.~Nirman Ganguly for his valuable insights.

\section*{Declarations}

\noindent\textbf{Author contributions.}
All authors contributed equally to the manuscript.
All authors have read and approved the final manuscript.

\medskip
\noindent\textbf{Data availability.}
Synthetic qubit datasets were generated and analysed in this study.

\medskip
\noindent\textbf{Competing interests.}
The authors declare no competing interests.

%%%%%%%%%%%%%%%%%%%%%%%%%%%%%%%%%%%%%%%%%%%%%%%%%%
\bibliographystyle{unsrtnat}
\bibliography{main}
\end{document}